# Dual Role of Nb in Defect-Mediated Strength and Ductility of γ-TiAl Alloys

Zhiqiang Zhao[1,2], Siyao Shuang[1,3], Kepeng Ouyang[2], Maolin Yu[4], Junping Du[5], Liangli Chu[2], Xiaokai Chen[2], Shigenobu Ogata[5], Wanlin Guo*[2], Zhuhua Zhang*[2] and Yong-Wei Zhang*[1]

*1 Institute of High-Performance Computing (IHPC), Agency for Science, Technology and Research (A∗STAR), 1 Fusionopolis Way, #16-16 Connexis, Singapore, 138632, Singapore*

*2 State Key Laboratory of Mechanics and Control for Aerospace Structures, Key Laboratory for Intelligent Nano Materials and Devices of Ministry of Education, and Institute for Frontier Science, Nanjing University of Aeronautics and Astronautics, Nanjing 210016, China*

*3 Applied Mechanics and Structure Safety Key Laboratory of Sichuan Province, School of Mechanics and Aerospace, Southwest Jiaotong University, Chengdu 610031, China*

*4 State Key Laboratory of Structural Analysis, Optimization and CAE Software for Industrial Equipment, Department of Engineering Mechanics, Dalian University of Technology, Dalian 116023, China*

*5 Department of Mechanical Science and Bioengineering, Graduate School of Engineering Science, The University of Osaka, Osaka, 560-8531, Japan*

*Correspondence: wlguo@nuaa.edu.cn (Wanlin Guo); chuwazhang@nuaa.edu.cn (Zhuhua Zhang); zhangyw@a-star.edu.sg (Yong-Wei Zhang)

**Abstract**: The origin of the superior high-temperature strength of γ-TiAl with high Nb addition remains highly controversial, largely due to the unclear role of Nb atoms. Using large-scale hybrid Monte Carlo and molecular dynamics simulations with a self-developed neural network potential, we show that Nb atoms predominantly occupy Ti sites and form short-range order with neighboring Al atoms, but a non-negligible fraction also occupies Al sites ($Nb_{Al}$) and promotes the formation of antisite defects ($Ti_{Al}$). Both the $Nb_{Al}$ and $Ti_{Al}$ antisites exceptionally reduce stacking fault energies and facilitate deformation twinning, thereby enhancing plasticity. Meanwhile, these substitutional and antisite defects also increase the Peierls stress of both screw and edge dislocations, which hinders dislocation motion to cause pronounced solid-solution strengthening. This work provides mechanistic insights into the dual role of Nb in enhancing both strength and ductility in γ-TiAl and further offers guidance for defect and composition engineering in advanced alloy systems.

## INTRODUCTION

γ-TiAl alloys uniquely combine low density, high specific strength, and superior creep and oxidation resistance[1, 2]. Their elastic modulus and creep resistance are comparable to those of Ni-based superalloys but with only about half the density, making them attractive candidates for replacing Ni-based alloys in aerospace engines operating at 650-800 °C, which enables significant weight reduction[3, 4, 5]. However, their limited room-temperature ductility and low fracture toughness, along with microstructural degradation during long-term high-temperature service, continue to hinder widespread application[2]. To address these issues, extensive efforts have focused on microstructural optimization and alloying strategies. In particular, high-Nb γ-TiAl alloys containing 5-10 at.% Nb exhibit markedly improved ordering temperatures and high-temperature mechanical properties[6, 7, 8], achieving increases of 60-100 °C in service temperature and improvements of 300-500 MPa in yield strength compared with conventional TiAl alloys. The development of polysynthetic twinned high-Nb TiAl single crystals[9] further pushed the service temperature above 900 °C, while enhancing room-temperature ductility over 6% and achieving yield strength beyond 700 MPa, demonstrating an exceptional balance of strength and ductility[4].

High-Nb alloying significantly enhances the mechanical properties of γ-TiAl, especially for Ti-rich alloys, yet its underlying strengthening mechanism remains highly debated. Paul *et al.*[10] attributed the improved strength to low Al content and associated microstructural changes, arguing that Nb plays a negligible role. However, the alloy samples used in their study exhibited distinct microstructures, and such pronounced differences obscure the intrinsic strengthening effect of Nb. To decouple composition from microstructure, Zhang *et al.*[11] experimentally measured the yield strength and friction stress of high-Nb and Nb-free TiAl alloys with similar near-γ microstructures and attributed the enhanced strength to Nb-induced solid-solution strengthening, with Nb preferentially occupying Al sublattices in Al-lean alloys. To understand the solid-solution strengthening, Woodward *et al.*[12] adopted a continuum model to study the flow behavior of γ-TiAl and reported that $Al_{Ti}$ antisites (Al atoms occupying Ti sites) have a strong strengthening effect, whereas $Ti_{Al}$ antisites (Ti atoms occupying Al sites) and Nb solutes have minimal effect. This contrasts with the experimental finding, which showed that solid-solution strengthening of Nb is more pronounced in Ti-rich than in Al-rich alloys[11].

Fröbel and Appel proposed that[13], in off-stoichiometric γ-TiAl, antisite and vacancy defects may form point defect complexes that generate orientation-dependent local strain fields. These complexes are thought to migrate toward dislocations via so-called anti-structural bridges (ASBs)[13], thereby impeding dislocation motion and contributing to the strengthening of γ-TiAl. However, DFT calculations by Li *et al.*[14] revealed that ASBs are energetically unfavorable in Ti-rich alloys, while ASBs more readily form on the Al-rich alloys. Thus, the ASBs-mediated strengthening mechanisms are insufficient to explain the high strength of Ti-rich γ-TiAl. Their calculations further indicated that Nb prefers Al sublattices and forms short-range order (SRO)[14], which they proposed as the origin of high strength in Ti-rich TiAl-Nb alloys. This hypothesis, however, contradicts numerous experimental and theoretical studies consistently showing that Nb strongly favors Ti sublattices. Moreover, experiments have demonstrated a significant reduction in stacking fault energies with Nb addition[15, 16], with the superlattice intrinsic stacking fault energy decreasing from ~63 to 34 mJ/m$^2$ as Nb content increases from 1 to 10 at.%, compared to 97 mJ/m$^2$ in Nb-free alloys[15, 16]. Yet, DFT results indicated that Nb increases stacking fault energies when occupying Ti sites, but significantly lowers it when on Al sites[17, 18, 19]. This discrepancy between experimental observations and theoretical predictions highlights a fundamental knowledge gap. In particular, the effects of Nb site occupancy and associated point defects (including antisites) on defect energetics and plasticity in high-Nb γ-TiAl alloy remain poorly understood. A comprehensive reassessment of Nb's dual role—both as a strengthener and a defect modifier—is urgently needed to resolve these inconsistencies and establish a unified mechanistic picture of strengthening in high-Nb γ-TiAl alloys.

In this work, we systematically investigate the site occupancy of Nb in γ-TiAl and its strengthening mechanisms using a high-accuracy neural network potential (NNP). Hybrid Monte Carlo and molecular dynamics simulations reveal that Nb strongly prefers Ti sites (Nb$_{Ti}$) and tends to form SRO with neighboring Al atoms, while also partially occupying Al sites (Nb$_{Al}$) and forming antisite defects (Ti$_{Al}$ and Al$_{Ti}$). Systematic stacking fault energies (SFEs) calculations demonstrate that Nb$_{Ti}$ defects increase both stable and unstable SFEs, whereas Nb$_{Al}$ defects and Ti$_{Al}$ antisites significantly reduce them. Energy barrier analyses confirm that Nb$_{Al}$ and Ti$_{Al}$ more effectively promote plastic deformation by reducing slip energy barriers compared to Nb$_{Ti}$ and Al$_{Ti}$ defects. Compared to defect-free γ-TiAl, the presence of substitutional (Nb$_{Ti}$ and Nb$_{Al}$) and antisite defects markedly increases the Peierls stress of both screw and edge dislocations, indicating pronounced solid-solution strengthening effects across all slip configurations. These results suggest that the SFE reduction in high-Nb γ-TiAl stems primarily from Nb$_{Al}$ and antisite defects, rather than Ti-site substitution alone, offering mechanistic insights into the dual role of Nb in simultaneously enhancing strength and ductility via defect-mediated pathways.

## MODELS AND METHODS

### Neural network potential for Ti-Al-Nb alloys

We recently developed a general-purpose neural network potential (NNP) for the Ti-Al-Nb ternary system by integrating the neural evolution potential framework[20] with an active learning scheme, as published in [*Phys. Rev. B* **110**, 184115 (2024)][17]. All simulations in this study were performed using this NNP-TiAlNb model based on GPUMD[20, 21] and LAMMPS[22] packages. Trained on comprehensive first-principles datasets, this NNP model exhibits excellent accuracy in predicting a wide range of lattice, defect, and thermodynamic properties, including thermal expansion and melting behavior of TiAl alloys. Notably, the developed NNP model captures the key effects of Nb doping on stacking fault energies and formation energies, making it particularly well-suited for investigating the atomic-scale mechanisms underlying Nb-induced strengthening in γ-TiAl alloys. Leveraging its efficiency and scalability, this model enables large-scale MD simulations involving tens of millions of atoms with only a few GPU cards, thus providing a powerful platform for exploring complex defect-dislocation interactions and compositional effects in TiAl-based systems.

*Hybrid Monte Carlo and molecular dynamics simulations*

To study the site occupancy preference of Nb, we performed hybrid Monte Carlo and molecular dynamics (MCMD) simulations on Nb-doped γ-TiAl samples. The algorithm alternates between MD relaxation steps and MC swap attempts, during which randomly selected pairs of atoms of different species exchange their atomic identities (masses and velocities). Swap acceptance follows the standard Metropolis criterion based on energy changes:

$$P = \min\left\{1, \exp\left(-\frac{\Delta U}{k_B T}\right)\right\}. \tag{1}$$

Here, $\Delta U$ denotes the change in potential energy resulting from a trial swap. For each MC trial swap, $\Delta U$ is evaluated accordingly. A key innovation for MCMD implemented in GPUMD lies in the optimized energy calculation scheme[23, 24]. Rather than recomputing the entire system's energy for each swap attempt, the short-range nature of interatomic potentials is exploited to update only local atomic environments. Each trial modifies at most ~4$M$ atomic interactions (where $M$ represents the average coordination number), making the MC phase computationally negligible compared to MD steps. This approach is efficiently implemented in the GPU-accelerated GPUMD framework, ensuring minimal computational overhead from the MC phase. This combined strategy is particularly well-suited for studying temperature-dependent site occupancy in complex alloy systems.

*Characterization of short-range ordering*

The Warren-Cowley parameter (WCP)[25] was used to characterize the SRO in the 1st-nearest neighbor shell of Nb-doped γ-TiAl samples:

$$\mathrm{WCP}_{mn} = 1 - Z_{mn} / (\chi_n Z_m), \tag{2}$$

where $\mathrm{WCP}_{mn}$ describes the degree of SRO between atomic species $m$ and $n$. It is defined based on $Z_{mn}$, the number of $n$-type atoms among the first-nearest neighbors of the $m$-type atom; $Z_m$, the total number of first-nearest neighbors of the m-type atom, and $\chi_n$, the atomic fraction of n-type atoms in the system. The value of $WCP_{mn}$ ranges from -2.0 to 1.0. A value close to -2.0 indicates a strong attractive tendency between $m$ and $n$-type atoms, whereas a value near 1.0 suggests strong repulsion. When $WCP_{mn}$ is close to zero, the $m$-$n$ type pairs are considered to be randomly distributed (no SRO).

*General stacking fault energies calculations*

General stacking fault energies (GSFEs) on the (111) plane were calculated at 0 K using a supercell containing 19,200 atoms, with dimensions of 90.45 × 98.82 × 34.83 Å³, as shown in Figure S1. Periodic boundary conditions were applied along the $x$ and $y$ directions, while a free surface was imposed along the $z$ direction. Prior to GSFE calculations, both the pristine and defect-containing γ-TiAl models were fully relaxed via energy minimization, during which all six pressure components were released. This procedure allows the simulation cell to naturally accommodate the intrinsic tetragonal distortion of the L1₀ γ-TiAl and fully captures the local lattice relaxation induced by substitutional and antisite defects. As a result, the computed GSFEs more accurately reflect the true anisotropic structural response of the γ-TiAl phase.

The relaxed models were then used for GSFE calculations using the LAMMPS package. In these simulations, the bottom half of the supercell was fully fixed, while the top half was incrementally displaced within the $x$-$y$ plane. During each displacement step, atoms in the top half were allowed to relax only along the $z$-direction, and the total energy of the system was minimized to obtain the corresponding GSFEs, $\gamma(\vec{u})$, as defined by the following equation:

$$\gamma(\vec{u}) = \frac{E_{fault}(\vec{u}) - E_0}{A}, \tag{3}$$

where $E_{fault}(\vec{u})$ denotes the total energy of the system after applying the relative shift along the fault plane, $E_0$ represents the total energy of the perfect crystal, and $A$ is the area of the fault plane.

*Peierls stress calculations*

To estimate the role of substitutional and antisite defects in strengthening γ-TiAl, we calculated the Peierls stress of both screw and edge dislocations. Figure S2a shows the initial screw-dislocation model with dimensions of approximately 347 × 195 × 101 Å³ and crystal orientations of $x$ = [11-2], $y$ = [111], and $z$ = [1-10]. A single 1/2[1-10] screw dislocation was introduced at the center using the anisotropic dislocation displacement field, resulting in a model system of 423,360 atoms. A displacement offset of $b$/2 was applied along the $z$ direction to maintain a single-dislocation configuration under periodic boundaries[26, 27]. For the edge-dislocation model (Figure S2b), the system contains 387,144 atoms with dimensions of ~336 × 210 × 90 Å³ and orientations of $x = [1\bar{1}0]$, $y = [111]$, and $z = [11\bar{2}]$. A single $1/2[1\bar{1}0]$ edge dislocation was introduced at the center by removing a half-plane below the glide plane. For all dislocation models, periodic boundary conditions were applied along the $x$ and $z$ directions, while free boundary condition was used in the $y$ direction.

To eliminate boundary-induced residual stress, each initial model was fully relaxed after dislocation insertion until all system-averaged stress components converged to zero (Figure S2c). To drive screw (edge) dislocations motion along the $[11\bar{2}]$ ($[1\bar{1}0]$) direction on the (111) plane, the bottom 10 Å of atoms were fixed while the top 10 Å was subjected to a constant velocity of $1\times10^{-5}$ Å/fs ($\sim5\times10^{7}$/s) along the $z$-direction ($x$-direction). The Peierls stress was determined by incrementally applying shear strain and performing energy minimizations at each step to identify the minimum stress required to initiate stable dislocation glide. To examine the influence of defect type on dislocation slip, four types of defects, including $Nb_{Ti}$, $Nb_{Al}$, $Al_{Ti}$, and $Ti_{Al}$, were introduced at a concentration of 8 at.%.

## RESULTS AND DISCUSSIONS

### Site occupation of Nb in γ-TiAl

γ-TiAl is an ordered L1$_0$ intermetallic compound with a face-centered tetragonal structure, where Ti and Al atomic layers are alternately arranged along the [001] direction (see inset in Figure 1a). The NNP-predicted lattice parameters of γ-TiAl are $a = 3.986$ Å, $c = 4.084$ Å, $c/a = 1.024$, showing excellent agreement with experimental values ($a = 3.988$ Å, $c = 4.067$ Å)[28]. The deviation of the axis ratio ($c/a$) from unity results in crystallographic anisotropy, as evidenced by differences in the lengths of the [001], [100], and [010] directions. In contrast to face-centered cubic metals, which possess 12 slip systems, the γ phase has only four, significantly limiting its plastic deformation capability and resulting in inherently brittle mechanical behavior[3].

Alloying with Nb is an effective approach to improve the mechanical properties of γ-TiAl. Understanding the site occupancy of Nb in the γ phase is crucial for revealing the underlying strengthening mechanisms of alloying effects. DFT calculations based on formation energy or enthalpy have suggested that Nb energetically prefers Ti sites[29, 30, 31], but these predictions often neglect the influence of temperature and alloy composition. As an alternative, thermodynamic statistical-mechanical models have been developed for studying site occupancy. Jiang *et al.*[32] used the Wagner-Schottky model and reported that Nb site preference is both composition- and temperature-dependent: at 0 K, Nb prefers Ti sites in Al-rich TiAl alloys and Al sites in Ti-rich TiAl alloys; while at high temperatures, entropy drives Nb to preferentially occupy Ti sites regardless of composition. Diao *et al.*[33] further incorporated defect interactions into the thermodynamic model and showed that while site occupancy remains composition-dependent at low temperatures, Nb atoms predominantly occupy Ti sublattices at high temperatures in both Ti-rich and Al-rich TiAl alloys. However, these empirical thermodynamic models rely on simplifying assumptions, which may hinder their predictive accuracy.

In contrast, the NNP-driven hybrid Monte Carlo and molecular dynamics scheme provides a powerful and efficient approach for investigating site occupancy behavior, as it enables large-scale sampling of thermally activated atomic configurations at realistic temperatures and compositions while retaining *ab initio* accuracy. In this part, we use NNP-driven MCMD simulations to examine the temperature- and composition-dependent site occupancy behavior of Nb in γ-TiAl. We first perform MCMD simulations at 900 K starting from an initial Nb-doped model, in which 8 at.% of Ti and Al atoms are randomly substituted by Nb atoms. As shown in Figure 1a, the potential energy of the model system decreases rapidly during the early stages of the MCMD simulation and gradually converges, indicating enhanced thermodynamic stability after structural relaxation. Figures 1b and 1c illustrate the distribution of Nb atoms before and after MCMD, respectively. Initially, Nb atoms are randomly distributed across both Ti and Al sublattices (Figure 1b). After MCMD optimization (Figure 1c), however, the vast majority of Nb atoms occupy Ti sites, with only a few remaining on Al sites, suggesting a clear site preference.

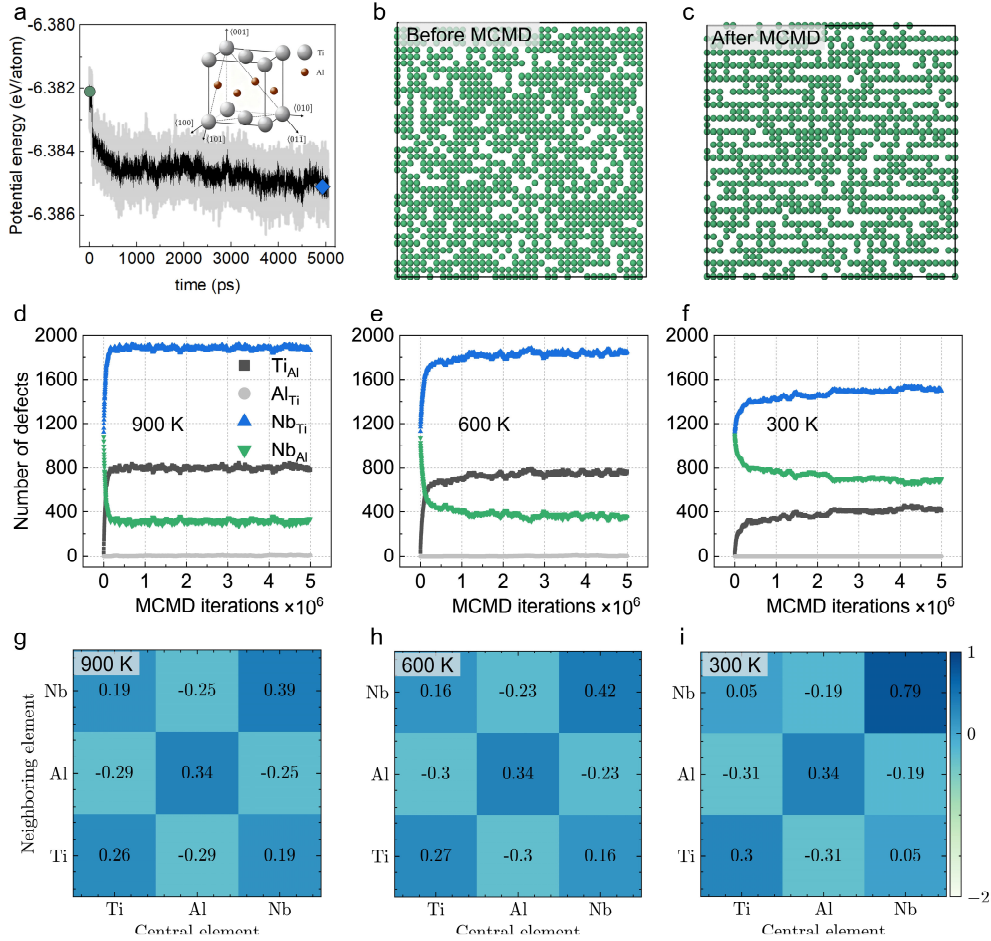

**Figure 1. MCMD simulations for investigating the site occupation of Nb in γ-TiAl. a** Potential

energy per atom as a function of time during the MCMD process at 900 K. The inset shows the crystal structure of γ-TiAl. **b-c** Distribution of Nb atoms in γ-TiAl lattice before (**b**) and after (**c**) MCMD simulations. For clarity, Ti and Al atoms are not shown. **d-f** Evolution of defect numbers as a function of MCMD iterations at 900 (**d**), 600 (**e**), and 300 K (**f**). **g-i** Table of Warren-Cowley parameters for MCMD-relaxed configurations at 900 (**g**), 600 (**h**), and 300 K (**i**).

To quantify the defect evolution induced by Nb doping, we track site occupancy statistics throughout the MCMD simulations from 900 to 300 K, as shown in Figure 1d-f. Four types of point defects are identified: two substitutional defects—$Nb_{Ti}$ and $Nb_{Al}$, where Nb occupies Ti and Al sites, respectively; two antisite defects—$Ti_{Al}$ (Ti atoms occupying Al sites) and $Al_{Ti}$ (Al atoms occupying Ti sites). Initially, only the $Nb_{Ti}$ and $Nb_{Al}$ defects are present in nearly equal numbers, reflecting the unbiased nature of the stochastic Nb substitution. As the MCMD simulation proceeds, the number of $Nb_{Ti}$ and $Ti_{Al}$ defects increases significantly, while $Nb_{Al}$ defects decrease sharply. After $5 \times 10^6$ MCMD steps, $Nb_{Ti}$ becomes the overwhelmingly dominant defect species, followed by $Ti_{Al}$ and $Nb_{Al}$, with $Al_{Ti}$ defects remaining scarce. At high temperatures of 900 (Figure 1d) and 600 K (Figure 1e), the defect populations consistently follow the trend: $Nb_{Ti} \gg Ti_{Al} > Nb_{Al} \gg Al_{Ti}$, indicating that the system approaches a thermodynamically favored distribution. This trend arises because Nb exhibits a strong preference for the Ti sublattice, driven by the lower formation energy of $Nb_{Ti}$ compared with $Nb_{Al}$[17]. Consequently, Nb atoms tend to migrate from Al to Ti sites through defect-conversion mechanisms such as:

$$Nb_{Al} + Ti_{Ti} \rightarrow Nb_{Ti} + Ti_{Al}. \tag{4}$$

This defect-conversion mechanism explains the concurrent increase of $Nb_{Ti}$ and $Ti_{Al}$, as well as the sharp reduction in $Nb_{Al}$. For example, at 900K (Figure 1d), $Nb_{Ti}$ increases from ~1100 to ~1900, $Nb_{Al}$ declines to ~ 300, and $Ti_{Al}$ grows from 0 to ~800. In contrast, at 300 K (Figure 1f), the defect populations change to $Nb_{Ti} \gg Nb_{Al} > Ti_{Al} \gg Al_{Ti}$. At such a low temperature, atomic diffusion is drastically suppressed, and the long-range or multi-atom cooperative movements required for Nb migration from Al to Ti sites rarely occur. As a result, although $Nb_{Ti}$ remains the most abundant defect species, a substantially larger fraction of $Nb_{Al}$ defects persists at 300 K compared with higher temperatures, reflecting a kinetically frozen, nonequilibrium configuration. Meanwhile, it is noted that at 300 K the number of $Al_{Ti}$ antisites is nearly zero (Figure 1f), while at 900 K the number of $Al_{Ti}$ defects increases to several tens (Figure 1d), suggesting an enhanced degree of sublattice disorder at high temperature. Overall, these results demonstrate a strong thermodynamic preference for Nb to occupy Ti sites, accompanied by the concurrent formation of $Ti_{Al}$ antisite defects when sufficient thermal activation is available.

These trends in site occupation are further revealed by the Warren-Cowley parameters obtained from atomic configurations before (Figure S3) and after the MCMD optimization (Figure 1g-i). At 900 K, the WCP value of Nb-Nb pairs increases from -0.01 (Figure S3) to 0.39 (Figure 1g) after MCMD optimization, indicating a strong mutual repulsion between Nb atoms and a tendency to avoid forming Nb-Nb nearest-neighbor pairs. The negative WCP of -0.29 for Ti-Al pairs (Figure 1g) reveals a strong tendency toward heteroatomic ordering, consistent with the long-range ordering of γ-TiAl. Meanwhile, the Nb-Al pairs also exhibit negative values of -0.25, implying a favorable chemical affinity between Nb and Al atoms in their nearest-neighbor environment. This is consistent with Nb primarily occupying Ti sublattice sites (Figure 1d), which are naturally surrounded by Al atoms in the ordered $L1_0$ lattice. In addition, the second-nearest-neighbor WCP for Nb-Al gives a large positive value of 0.82 (Figure S4), further supporting the formation of short-range ordering Nb-Al pairs. In contrast, the positive Nb-Ti WCP of 0.19 (Figure 1g) indicates repulsion and an unfavorable Nb-Ti nearest-neighbor environment (Nb occupying Al sites). As the temperature decreases from 900 to 300 K, the WCP of Nb-Al decreases from -0.25 (Figure 1g) to -0.19 (Figure 1i), which indicates that the number of $Nb_{Ti}$ gradually decreases due to limited atomic diffusions (Figure 1d-f). Overall, the WCP analysis further confirms a strong thermodynamic preference for Nb to occupy Ti sites and form Nb-Al short-range order at high temperatures.

### *Several factors on site occupation*

We also investigate the site preference of Nb atoms in γ-TiAl across a range of doping concentrations at 900 K. The initial Nb-doped γ models are constructed by randomly substituting either Ti or Al atoms with Nb atoms at concentrations ranging from 2 to 10 at.% in increments of 2 at.%. Figures 2a-c show representative defect evolution at 2, 6, and 10 at.% Nb doping, while results for intermediate concentrations are provided in Figure S5. Across all concentrations studied, the number of each defect type consistently followed the order: $Nb_{Ti} \gg Ti_{Al} > Nb_{Al} \gg Al_{Ti}$, only with minor variations in their relative proportions. By normalizing the number of $Nb_{Ti}$ and $Nb_{Al}$ defects to the total number of Nb atoms, we can quantify their site occupancy ratios as a function of doping concentration (Figure 2d). As the Nb content increases from 2 to 10 at.%, the ratio of $Nb_{Ti}$ decreases from ~95.5% to 84%, whereas $Nb_{Al}$ increases from ~4.5% to 16%. Meanwhile, a clear upward trend in $Ti_{Al}$ antisites is observed with increasing Nb concentration (Figure 2a-c). Thus, it can be expected that a large fraction of $Nb_{Al}$ and $Ti_{Al}$ exists in high-Nb TiAl alloys.

To further elucidate the effects of temperature and stoichiometry on Nb site preference, we perform MCMD simulations using γ-TiAl models doped with 8 at.% Nb. MCMD simulations were performed from 900 to 300 K (Figure 1d-f) in the interval of 150 K (see Figure S6 for the results at intermediate temperatures). As the temperature increases from 300 to 900 K, the fraction of $Nb_{Ti}$ increases from ~68% to 86%, while $Nb_{Al}$ decreases from 32% to 14% (Figure 2e), indicating a stronger thermodynamic preference for Nb occupying Ti sites at high temperature. The influence of stoichiometry is further examined at 900 K by comparing three compositions: Ti-45Al-8Nb (Ti-rich), Ti-46Al-8Nb (stoichiometric), and Ti-47Al-8Nb (Al-rich). All models exhibit Nb$_{Ti}$-dominated behavior (Figure S7), though with notable differences. The Ti-rich sample shows the lowest Nb$_{Ti}$ fraction (~82%) and highest Nb$_{Al}$ (~18%), whereas the Al-rich sample displays the opposite trend, with Nb$_{Ti}$ ~90% and Nb$_{Al}$ ~10% (Figure 2f). Ti-rich alloys also contained the most Ti$_{Al}$ antisites (Figure S7), underscoring that an excess of Ti increases off-site disorder, while Al excess suppresses it. These observations reveal that both elevated temperature and Al-rich chemistry favor Nb occupancy on Ti sites, and that deviations from the ideal Ti:Al ratio modulate defect formation, with implications for targeted defect engineering in Nb-doped γ-TiAl alloys.

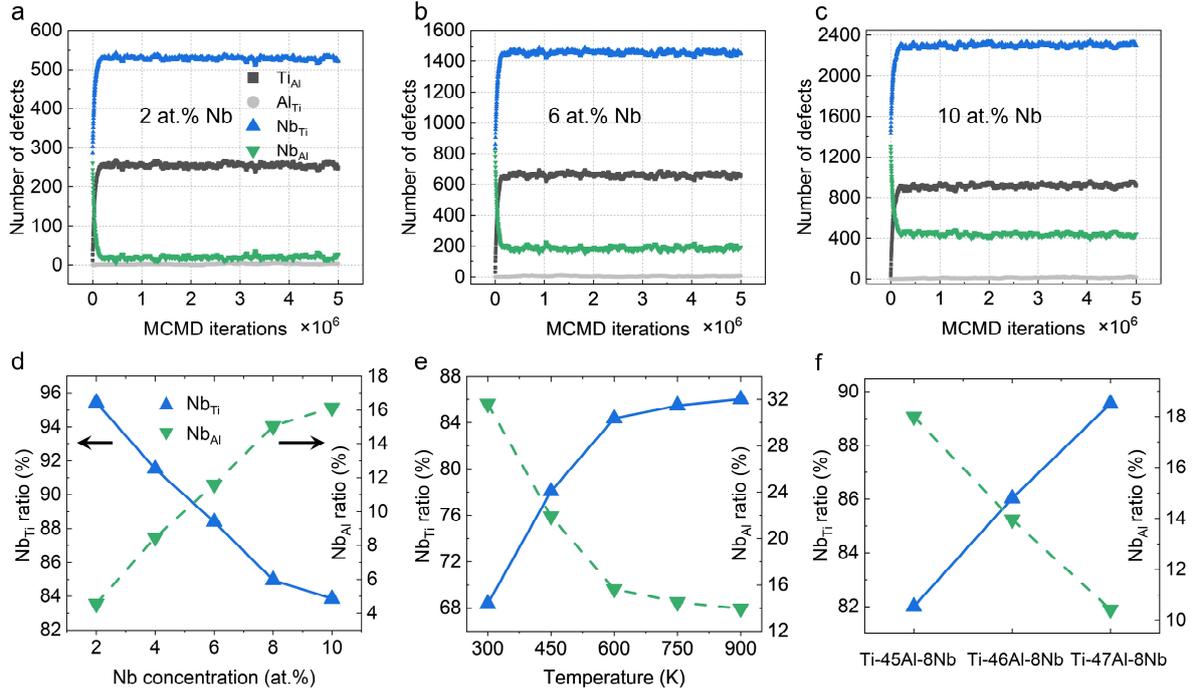

**Figure 2. Concentration-, temperature-, and composition-dependent site occupation of Nb in γ-TiAl. a-c** Evolution of defect numbers during MCMD iterations for Nb doping concentrations of 2 at.% (**a**), 6 at.% (**b**), and 10 at.% (**c**). **d** Ratio of Nb$_{Ti}$ to Nb$_{Al}$ as a function of Nb doping concentration. **e** Ratio of Nb$_{Ti}$ to Nb$_{Al}$ as a function of temperature. **f** Ratio of Nb$_{Ti}$ to Nb$_{Al}$ for different compositions of TiAl-Nb alloys.

### *Comparison with experimental results on site occupation*

In this part, we have systematically investigated the site occupation of Nb in γ-TiAl under thermodynamic conditions using the NNP-driven MCMD method. Across a range of doping concentrations, temperatures, and chemical stoichiometries, Nb consistently exhibits a very strong preference for substituting Ti sites. The defect population in all simulation cases follows the order: Nb$_{Ti}$ ≫ Ti$_{Al}$ > Nb$_{Al}$ ≫ Al$_{Ti}$ , indicating Nb$_{Ti}$ is energetically most favorable. Our results agree well with experimental studies using atom location channeling enhanced microanalysis (ALCHEMI), X-ray scattering, and field-ion microscope (FIM), which consistently report Nb occupying Ti sublattices[34, 35, 36, 37, 38, 39], regardless of alloy compositions or heat treatment, as summarized by Hu *et al.*[33] (Table S1). Notably, Mohandas *et al.*[40] observed that ~11% of Nb atoms occupy Al sites in Ti-54.0Al-2Nb (Table S1). Consistently, our MCMD simulations also predict a minor but non-negligible population of Nb$_{Al}$ defects.

It is important to note that existing experimental studies have primarily focused on TiAl alloys with relatively low Nb concentrations (<5 at.%). Our simulations indicate that the probability of Nb atoms occupying Al sublattices increases with Nb content, suggesting that Nb$_{Al}$ defects become more numerous in high-Nb TiAl alloys than previously recognized. Therefore, while N$_{Ti}$ is the overwhelmingly dominant defect species under equilibrium conditions, the formation of Nb$_{Al}$ under certain thermodynamic and kinetic pathways is plausible, and supported by both our MCMD simulations and selected experimental observations. These findings highlight the importance of considering both equilibrium thermodynamics and kinetic constraints when interpreting site occupancy in complex intermetallic systems.

### *Intrinsic plastic deformation of γ-TiAl*

Building upon the site preference of Nb, we further investigate the effect of Nb alloying on the plastic deformation behavior of γ-TiAl. As a foundation, we first analyze the intrinsic deformation mechanisms of the defect-free γ-TiAl by computing the generalized stacking fault energies (GSFEs) using the NNP model. In the L1$_0$ γ-TiAl, plastic deformation is governed by slip along the close-packed {111} planes. To quantify the energetics of such processes, GSFEs curves on the (111) planes are calculated using a supercell containing 19,200 atoms, with dimensions of 90.45 × 98.82 × 34.83 Å$^3$ (see Models and Methods section for more details).

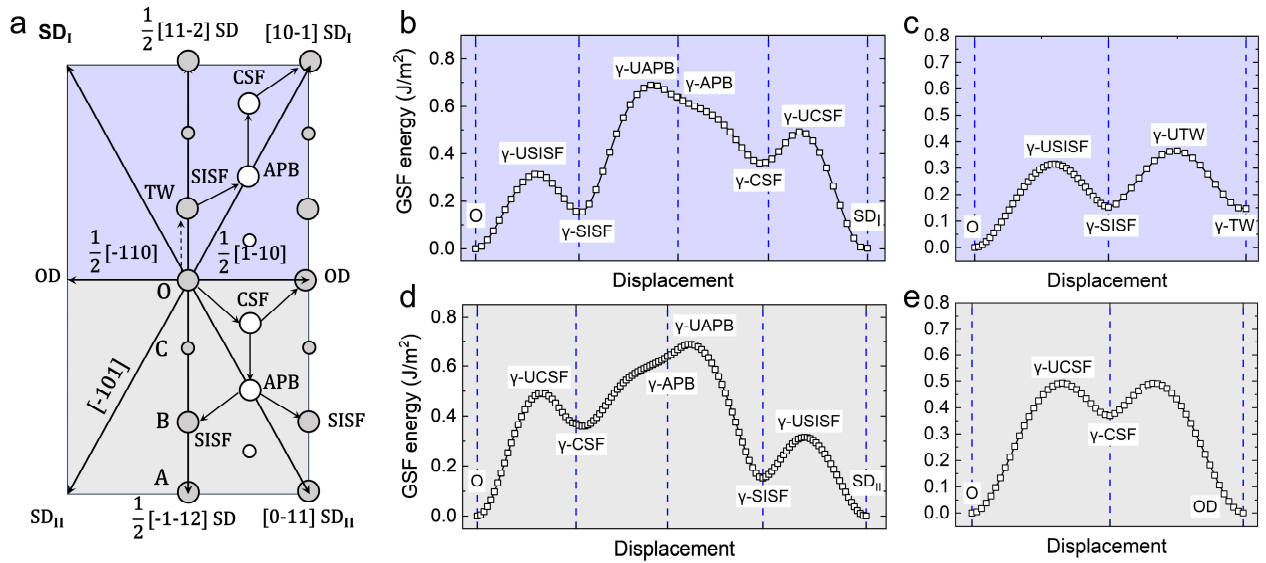

**Figure 3. Plastic deformation of γ-TiAl. a** Schematic for the slip and twinning systems on (111) planes of γ-TiAl, together with Burgers vectors for ordinary dislocations (OD), twinning (TW), and superlattice dislocations (SD). **b-c** Generalized stacking fault energies curves for the SD$_I$ (**b**) and TW (**c**) deformation modes, both of which are initiated by the superlattice intrinsic stacking

fault (SISF) partial. **d-e** Generalized stacking fault energies curves for the SD$_{II}$ (**d**) and OD (**e**) deformation modes, both of which are initiated by the complex stacking fault (CSF) partial.

γ-TiAl has three primary deformation modes: ordinary dislocations (OD), twinning (TW), and superlattice dislocations (SD). The characteristic Burgers vectors are 1/2[1-10] for OD and 1/6[11-2] for TW. For SD modes, the corresponding Burgers vectors are [10-1] for SD$_I$, [0-11] for SD$_{II}$, and 1/2[11-2] for SD. Following the work of Levente Vitos *et al.*[41], Figure 3a schematically illustrates the corresponding deformation pathways on the (111) plane. The schematic highlights the energetically favorable dislocation dissociation paths and their associated Burgers vectors. In this depiction, Ti and Al atoms are represented by gray and white circles, respectively. The difference in circle size denotes the three adjacent (111) atomic layers, labeled A, B, and C. Regions shaded in purple and gray correspond to deformation modes initiated by the superlattice intrinsic stacking fault (SISF) and the complex stacking fault (CSF), respectively. For a clear illustration, the corresponding GSFE surface of the {111} plane in γ-TiAl is shown in Figure S8.

The dislocation dissociation processes associated with each deformation mode are summarized as follows. For the OD deformation mode, the two-fold dislocation dissociation is described as:

$$\text{OD}: 1/2[1\bar{1}0] \rightarrow 1/6[1\bar{2}1] + \text{CSF} + 1/6[2\bar{1}\bar{1}]. \tag{5}$$

The OD mode involves slip along the (111) plane in the [1-21] direction by a displacement of 1/6[1-21] to form the CSF, followed by an additional slip of 1/6[2-1-1]. The TW deformation mode proceeds via two successive partial dislocation slips along the [11-2] direction. The initial slip along 1/6[11-2] direction generates the SISF, followed by an equivalent 1/6[11-2] slip on the adjacent (111) atomic plane, resulting in the formation of a three-layer twin structure. The SD deformation mode can be further classified into SD$_I$ and SD$_{II}$ according to the difference in the leading partial dislocations. For the SD$_I$ mode, where the SISF is the leading partial, the fourfold dissociation can be expressed as:

$$\text{SD}_I: [10\bar{1}] \rightarrow 1/6[1\bar{1}\bar{2}] + \text{SISF} + 1/6[2\bar{1}\bar{1}] + \text{APB} + 1/6[11\bar{2}] + \text{CSF} + 1/6[2\bar{1}\bar{1}]. \tag{6}$$

For the SD$_{II}$, where the CSF is the leading partial, the dissociation may occur as follows:

$$\text{SD}_{II}: [0\bar{1}1] \rightarrow 1/6[1\bar{2}1] + \text{CSF} + 1/6[\bar{1}\bar{1}2] + \text{APB} + 1/6[1\bar{2}1] + \text{SISF} + 1/6[\bar{1}\bar{1}2]. \tag{7}$$

Figures 3b-c present the GSFE curves corresponding to SD$_I$ and TW modes, where the SISF acts as the leading partial. Figures 3d-e display the GSFE profiles for SD$_{II}$ and OD, in which the CSF serves as the leading partial. These energy landscapes reveal key stable stacking fault energies ($\gamma_{SFEs}$), including $\gamma_{SISF}$, $\gamma_{CSF}$, $\gamma_{TW}$, and $\gamma_{APB}$, along with the associated unstable stacking fault energies ($\gamma_{USFEs}$), such as $\gamma_{USISF}$, $\gamma_{UCSF}$, $\gamma_{UTW}$, and $\gamma_{UAPB}$. For the SD$_I$ deformation mode (Figure 3b), three energy barriers are identified: $\gamma_{USISF} < \gamma_{UCSF} < \gamma_{UAPB}$, and three corresponding stable stacking

faults follow the order: $\gamma_{SISF} < \gamma_{CSF} < \gamma_{APB}$, which is consistent with experimental observations[42, 43]. The TW mode involves two barriers of $\gamma_{USISF}$ and $\gamma_{UTW}$ (Figure 3c), with $\gamma_{UTW}$ slightly higher, and features two stable stacking faults: $\gamma_{SISF}$ and $\gamma_{TW}$. The SD$_{II}$ mode (Figure 3d) also involves three $\gamma_{USFEs}$ and three $\gamma_{SFEs}$, similar to the SD$_I$ mode. Specifically, $\gamma_{CSF}$ and $\gamma_{APB}$ are located between $\gamma_{UCSF}$ and $\gamma_{UAPB}$, while $\gamma_{SISF}$ lies between $\gamma_{UAPB}$ and $\gamma_{USISF}$. In the OD mode (Figure 3e), two identical $\gamma_{UCSF}$ barriers are observed, with the $\gamma_{CSF}$ situated in between. These observations collectively highlight the intricate relationship between dislocation dissociation behavior and the underlying GSFEs landscape, offering critical insights into the deformation mechanisms of TiAl alloys.

### *Effect of defects on plastic deformation*

Upon establishing the fundamental deformation modes of γ-TiAl, we further investigate the influence of defect types on stacking fault energies. As shown in Figure 4a, all stable SFEs increase with increasing Nb$_{Ti}$ content, while they decrease markedly with increasing Nb$_{Al}$ content. Specifically, as the Nb$_{Ti}$ concentration increases from 0 to 10 at.%, $\gamma_{SISF}$ rises from ~152 to 180 mJ/m$^2$, and $\gamma_{TW}$ exhibits a similar upward trend. In contrast, $\gamma_{APB}$ and $\gamma_{CSF}$ show more pronounced increases. When Nb substitutes for Al sites, the trend reverses: $\gamma_{SISF}$ decreases from ~152 to 80 mJ/m$^2$, and $\gamma_{TW}$ drops from ~144 to 82 mJ/m$^2$, with $\gamma_{APB}$ and $\gamma_{CSF}$ showing even steeper declines. Figure 4b shows that the $\gamma_{USISF}$ and $\gamma_{UCSF}$ are nearly invariant with increasing Nb$_{Ti}$ concentration, whereas $\gamma_{UTW}$ and $\gamma_{UAPB}$ exhibit slight increases. In contrast, all unstable SFEs decrease substantially with increasing Nb$_{Al}$ concentration, suggesting that Nb substituting for Al sites facilitates dislocation slip by lowering the associated energy barriers, thereby enhancing the plastic deformability of TiAl alloys.

Figures 4c and 4d illustrate the dependence of stable and unstable SFEs on antisite defects concentrations. In non-stoichiometric TiAl-based alloys, excess Ti or Al atoms are accommodated in the form of antisite defects (Ti$_{Al}$ or Al$_{Ti}$), which can also be regarded as a special type of alloying element. As shown in Figure 4c, all stable SFEs decrease with increasing Ti$_{Al}$ antisite content. For instance, as the Ti$_{Al}$ concentration increases from 0 to 10 at.%, $\gamma_{SISF}$ drops sharply from ~152 to 50 mJ/m$^2$. The presence of Al$_{Ti}$ antisite defects, on the other hand, reduces $\gamma_{APB}$ and $\gamma_{CSF}$ but increases $\gamma_{TW}$ and $\gamma_{SISF}$. Regarding the unstable SFEs, all four $\gamma_{USFEs}$ decrease with increasing Ti$_{Al}$ antisite content, again implying that Ti$_{Al}$ antisites potentially promote dislocation nucleation and glide by reducing the slip energy barriers. In Al-rich alloys, a similar trend is observed: $\gamma_{UCSF}$ and $\gamma_{UAPB}$ decrease with increasing Al$_{Ti}$ concentration, albeit to a lesser extent than in Ti-rich systems with Ti$_{Al}$ defects. Notably, $\gamma_{USISF}$ and $\gamma_{UTW}$ increase with higher Al$_{Ti}$ antisite concentrations, indicating that Al$_{Ti}$ antisites impede the slip of SISF and TW, while facilitating the slip of CSF and SD. The experimentally measured $\gamma_{SISF}$ values for Ti-54Al[44], Ti-49.6Al[45] and Ti-48.0Al[45] alloys are ~140, 98 and 67 mJ/m$^2$, respectively, confirming a decreasing trend with increasing Ti content (see green pentagrams in Figure 4c), in agreement with our theory predictions. It should be noted that experimental determinations of SFEs are subject to the limited resolution of TEM techniques, which can introduce uncertainties and lead to underestimations.

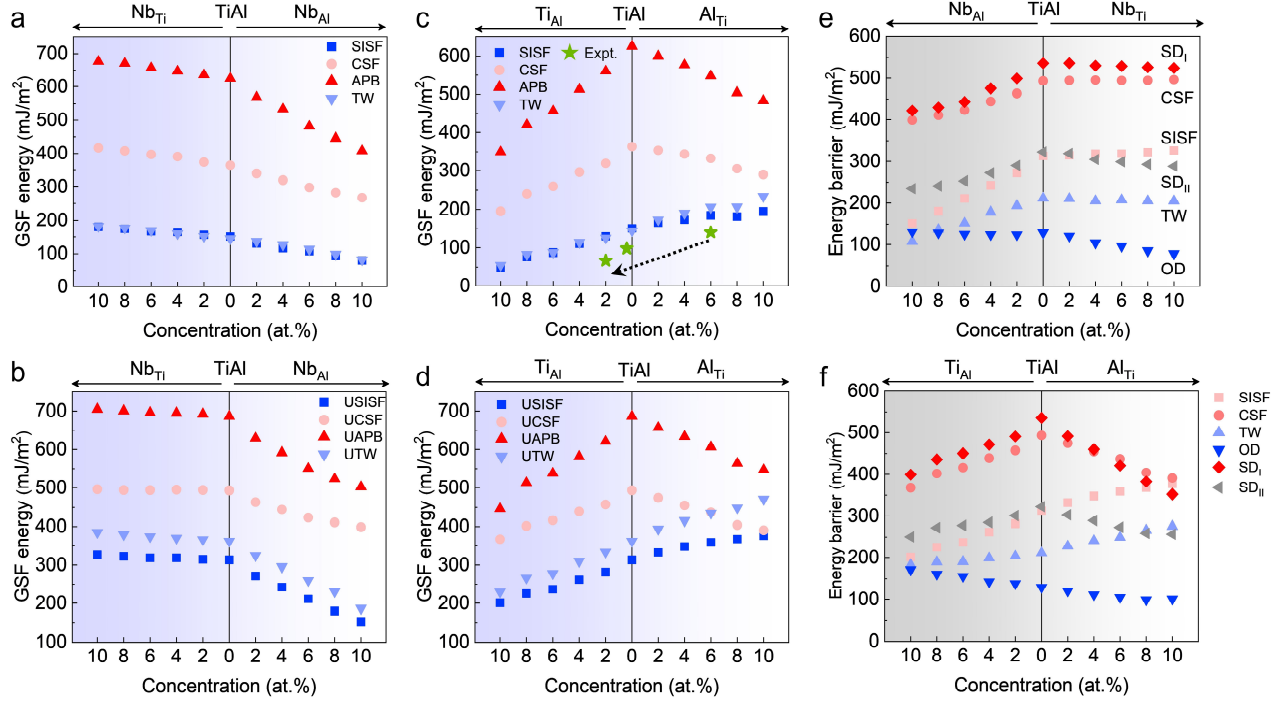

**Figure 4. Effect of defects on the stable and unstable stacking fault energies of γ-TiAl, as well as on energy barriers. a-b** Calculated stable (**a**) and unstable stacking energies (**b**) as a function of $Nb_{Ti}$ and $Nb_{Al}$ defect concentrations. **c-d** Calculated stable (**c**) and unstable stacking energies (**d**) as a function of $Ti_{Al}$ and $Al_{Ti}$ antisite defect concentrations. Green pentagrams represent the experimental data[44, 45]. **e-f** Calculated energy barriers as a function of substitutional (**e**) and antisite (**f**) defect concentrations, respectively.

In summary, our results demonstrate that Nb substituting at Ti sites increases stacking fault energies of γ-TiAl, whereas both $Nb_{Al}$ and $Ti_{Al}$ antisite defects lead to a significant reduction in SFEs. Hybrid MCMD simulations show that although $Nb_{Ti}$ remains dominant across all doping levels, the fraction of $Nb_{Al}$ increases with Nb concentration, accompanied by a notable increase in $Ti_{Al}$ antisite defects. Therefore, we propose that the combined presence of $Nb_{Al}$ and $Ti_{Al}$ defects is a key factor contributing to the experimentally observed SFE reduction in Ti-rich γ-TiAl with high Nb content.

### *Deformation energy barriers*

The slip energy barrier for each deformation mode is usually governed by the highest energy barrier encountered along its corresponding slip pathway (Figure 3). Specifically, the energy barriers (EBs) for TW, $SD_I$, OD, and $SD_{II}$ deformation modes[46] are defined as $\gamma_{UTW}$-$\gamma_{SISF}$, $\gamma_{UAPB}$-$\gamma_{SISF}$, $\gamma_{UCSF}$-$\gamma_{CSF}$, and $\gamma_{APB}$-$\gamma_{CSF}$, respectively, while the partial dislocations of SISF and CSF have barriers of $\gamma_{USISF}$ and $\gamma_{UCSF}$. Figures 4e and 4f reveal the evolution of these slip energy barriers with varying concentrations of different defects. Among the four deformation modes, $SD_I$ exhibits the highest energy barrier of ~536 mJ/m$^2$, indicating it is the most difficult mode to activate, consistent with its rare observation in experiments. In contrast, OD has the lowest energy barrier (~129 mJ/m$^2$), enabling easier dislocation propagation via this path.

When Nb substitutes for Al sites (Figure 4e), the EBs for all deformation modes decrease significantly with increasing Nb content, except for the OD mode, which remains nearly unchanged. This behavior indicates that $Nb_{Al}$ defects broadly facilitate slip and promote plastic deformation. In contrast, when Nb occupies Ti sites, the EBs of the TW mode remain largely unaffected, while the EBs associated with SISF and CSF exhibit a slight increase. The EBs of the other modes continue to decline (Figure 4e). These results suggest that $Nb_{Ti}$ exerts a potential suppressive effect on the formation of SISF and CSF partials. Although $Nb_{Ti}$ facilitates the slip of OD and $SD_{II}$ modes led by the CSF, the corresponding CSF energy barrier ($\gamma_{UCSF}$) is substantially higher than the SISF barrier ($\gamma_{USISF}$). Consequently, OD and $SD_{II}$ modes tend to exhibit higher strength and lower ductility than the TW mode led by the SISF partial. Therefore, Nb substitution at Al sites yields a more significant enhancement of plasticity in TiAl alloys, as it effectively reduces nearly all EBs, in contrast to Nb substitution at Ti sites.

Further analysis shows that antisite defects also significantly influence the slip behavior of the $\gamma$-TiAl, as illustrated in Figure 4f. $Ti_{Al}$ antisite defects significantly reduce the energy barriers of SISF and CSF partials and continuously lower the barrier of the TW mode, while increasing that of the OD mode. This trend indicates a strong promotion of twinning-dominated plasticity. In contrast, $Al_{Ti}$ antisites slightly increase the barriers of SISF and TW but reduce those of OD, CSF, $SD_I$, and $SD_{II}$. Notably, $Al_{Ti}$ defects (similar to $Nb_{Ti}$ defects) facilitate OD and $SD_{II}$ slip modes led by the CSF. However, because the CSF barrier remains much higher than that of the SISF, the activation of OD and $SD_{II}$ is still hindered, resulting in high strength and low ductility. Therefore, from the perspective of deformation twinning, the effect of $Al_{Ti}$ defects on plasticity improvement is less pronounced than that of $Ti_{Al}$ defects. These trends align with experimental findings by Hall *et al.*[47], where Ti$_{52}$Al$_{48}$ (Ti-rich) alloys exhibited a high density of deformation twins and superior ductility compared to Ti$_{48}$Al$_{52}$ (Al-rich).

### *Lattice distortion induced by defects*

Beyond their impact on plastic behavior, these point defects also induce notable lattice distortions, which can significantly affect the mechanical properties of $\gamma$-TiAl. We then employ the NNP model to further investigate how various defect types and concentrations influence the lattice constants of $\gamma$-TiAl. When Nb substitutes Ti atoms, the *c*-axis expands slightly while the *a*- axis remains nearly unchanged (Figures 5a-c), resulting in a slight increase of the $c/a$ ratio from ~1.02 to 1.025 (10 at.% $Nb_{Ti}$). This trend is in excellent agreement, both qualitatively and quantitatively, with experimental observations by Tetsui *et al.*[48] (Figure S9a), further validating the predictive accuracy of our NNP model. In contrast, the classical EAM-Farkas[49] potential incorrectly predicts a pronounced increase in the $a$-axis and a significant decrease in the $c$-axis with $Nb_{Ti}$ doping[50] (Figure S9b), contradicting experimental results and exposing the model's limitations in simulating Nb-containing TiAl systems.

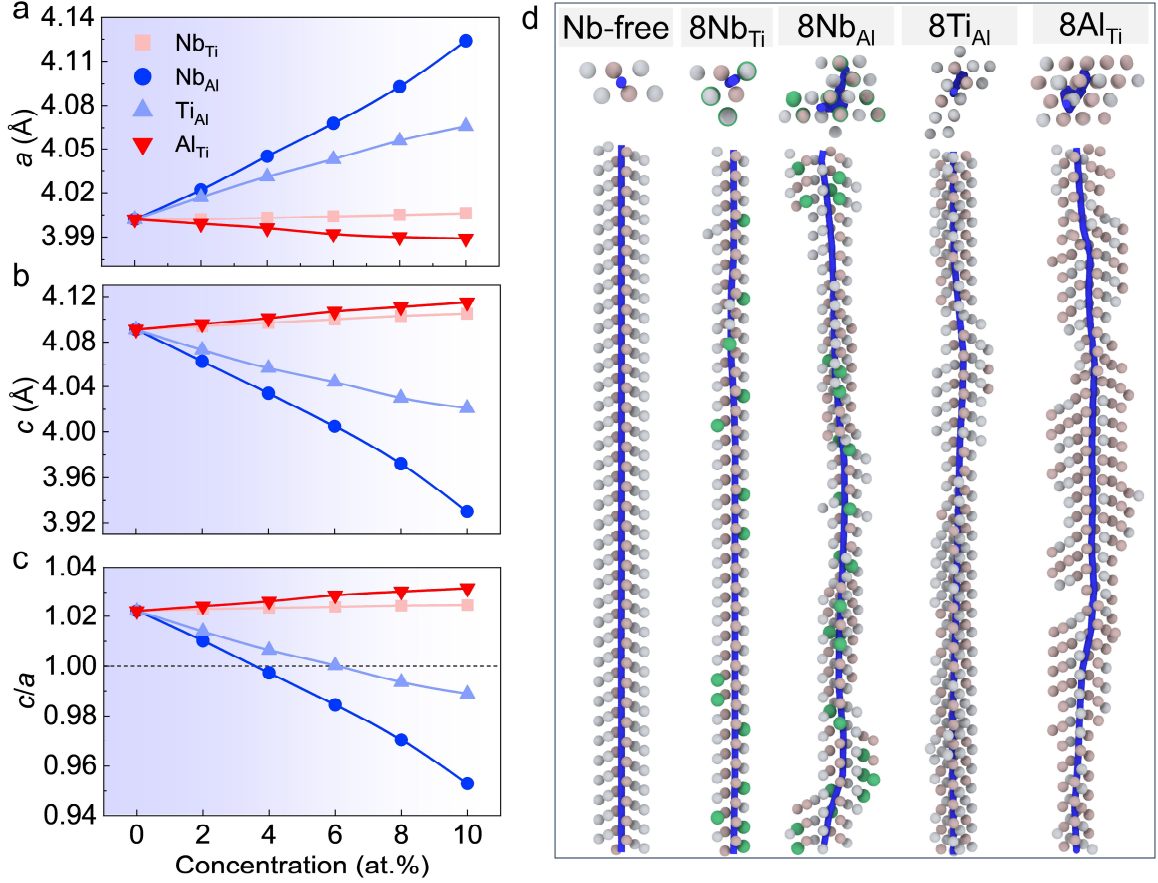

**Figure 5. Effect of defects on the lattice parameters and screw dislocation configuration of γ-TiAl. a-c** Lattice parameters $a$ and $c$, as well as the axis ratio $c/a$, as a function of defect concentrations. The simulation temperature for lattice parameter calculations is set to 300 K. **d** Relaxed screw dislocation core configurations under different defect types. Gray and brown spheres represent Ti and Al atoms, respectively, while green spheres represent Nb atoms. Blue lines indicate screw dislocation lines identified using the Dislocation Extraction Algorithm (DXA)[51] implemented in the OVITO code[52].

When Nb occupies Al sites, the lattice undergoes a stronger anisotropic distortion characterized by a substantial expansion of the $a$-axis (Figure 5a) and contraction of the $c$-axis (Figure 5b). This distortion, arising from the larger atomic size mismatch between Nb and Al (~2%) than between Nb and Ti (~0.6%), reduces the $c/a$ ratio from ~1.02 to 0.95 (Figure 5c). The ductility of γ-TiAl alloys is strongly correlated with the axial ratio. A reduction of $c/a$ toward unity increases lattice isotropy and diminishes the mobility contrast between ordinary and superdislocations, enabling more homogeneous plastic flow and thereby enhancing ductility[53]. In contrast, increasing $c/a$ enhances structural anisotropy, which strengthens the lattice but promotes brittle fracture. Consistently, Kawabata *et al.*[53] reported that lowering $c/a$ from 1.022 to ~1.012 increases the bend fracture strain of Ti-rich TiAl alloys from ~1.0% to over 2.5%, underscoring the beneficial role of a reduced axial ratio. Thus, $Nb_{Al}$ defects improve ductility within an appropriate concentration range by reducing the axial ratio. Meanwhile, at higher doping levels, they induce a stronger strengthening effect than $Nb_{Ti}$ defects by enhancing structural anisotropy.

For antisite defects, increasing the $Ti_{Al}$ antisite concentration leads to $a$-axis (Figure 5a) elongation and $c$-axis contraction (Figure 5b), decreasing the $c/a$ ratio to ~0.99 (10 at.% $Ti_{Al}$). Thus, $Ti_{Al}$ defects are favorable for plastic deformations in γ-TiAl. In contrast, $Al_{Ti}$ antisites cause $a$-axis contraction and $c$-axis elongation, increasing $c/a$ to ~1.03 (Figure 5c). These results indicate that $Al_{Ti}$ defects enhance lattice anisotropy and also increase the energy barriers for the twinning mode (Figure 4f), which strengthens the lattice by impeding dislocation motion but significantly increases the material's brittleness. These trends align well with the compression experiments by Tsujimoto *et al.*[54], which revealed that the plasticity of TiAl alloys is highly sensitive to their Al content: Al-rich alloys, typically containing $Al_{Ti}$ antisites, exhibit brittle behavior; whereas Ti-rich alloys with $Ti_{Al}$ antisites show enhanced ductility. Overall, within appropriate doping levels, both $Nb_{Al}$ and $Ti_{Al}$ defects can effectively improve the ductility of TiAl alloys, as also supported by the GSFEs in Figures 4e-f.

*Defect effects on dislocation motions*

Dislocation slip is a fundamental mechanism of plastic deformation in metals. The minimum shear stress required to move a single dislocation through a perfect lattice along a given slip plane, known as the Peierls stress[55, 56], is challenging to measure experimentally and typically assessed via atomic simulations. To elucidate the strengthening effect of Nb alloying, we perform shear simulations of γ-TiAl models containing either a screw or an edge dislocation (see Models and Methods for more details). We then systematically investigate the influence of different defect types on dislocation behavior and the resulting Peierls stress. Figure 5d depicts the relaxed configurations of screw dislocations at the zero-stress state. The dislocation in the Nb-free (pure γ-TiAl without defects) model is straight and undisturbed due to the absence of substitutional and antisite defects. In the 8 at.% $Nb_{Ti}$ model (8$Nb_{Ti}$), a slight curvature is observed along the dislocation line, whereas the 8 at.% $Nb_{Al}$ (8$Nb_{Al}$) model exhibits significant dislocation bowing, indicating a strong solid-solution strengthening effect when Nb occupies Al sites. Both $Ti_{Al}$ and Al$_{Ti}$ antisite defect models also show pronounced dislocation bending, further suggesting a substantial strengthening effect.

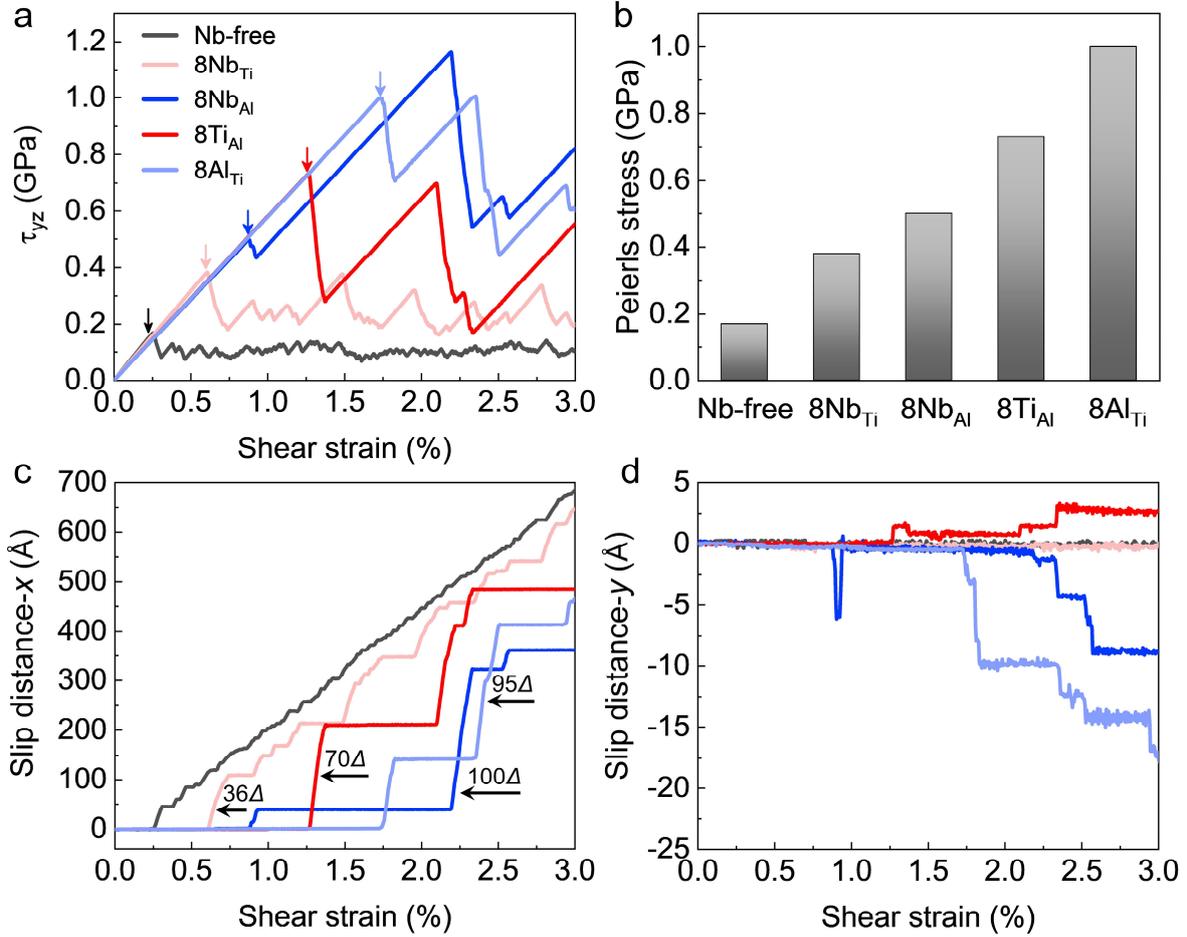

**Figure 6. Screw dislocation slips and mechanical responses of Nb-free (pure γ-TiAl) and defect-containing γ-TiAl models. a** Shear stress (τ$_{yz}$) as a function of shear strain for Nb-free and defect-containing models. **b** Calculated Peierls stress for the different models. **c-d** Slip distance of the screw dislocation along the *x*- (**c**) and *y*- (**d**) directions.

The obtained shear stress-strain curves in Figure 6a reveal that the Nb-free model shows a smooth, linear increase in stress, peaking at ~170 MPa, followed by mild fluctuations. In contrast, all four defect-containing models exhibit pronounced serrated flow behavior in stress-strain curves. The Peierls stress increases markedly in these models: ~380 MPa for Nb$_{Ti}$ (Figure 6b), more than double that of the Nb-free model (170 MPa), and ~500 MPa for the 8 at.% Nb$_{Al}$ model, indicating stronger strengthening when Nb replaces Al. For antisite defects, the Al$_{Ti}$ model yields the highest Peierls stress of ~1000 MPa, while the Ti$_{Al}$ model reaches ~700 MPa (Figure 6b). The above results indicate that both Nb substitution at Ti or Al sites, as well as the presence of antisite defects, exert a pronounced locking effect on dislocation motion. As shown in Figure 6c, the locking effect is further elucidated by the displacement-strain curves of dislocation motion along the *x*-direction (dislocation slip direction). Compared to the Nb-free model, the defect-containing systems exhibit longer jump distances following dislocation unlocking, accompanied by incubation times that scale approximately with the jump distance. This behavior highlights the intermittent nature of dislocation glide in the presence of solute atoms or antisite defects. The spacing between neighboring Peierls valleys on the {111} slip plane is $\Delta = a/\sqrt{2} \approx 2.81$ Å. Under the applied strain rate, maximum jumps of ~36$\Delta$ and 100$\Delta$ are observed in the 8Nb$_{Ti}$ and 8Nb$_{Al}$ models, respectively. Similarly, the Ti$_{Al}$ and Al$_{Ti}$ antisite models also exhibit jump distances of up to ~70$\Delta$ and 95$\Delta$, respectively. These extended jump events imply increased energy barriers that hinder continuous dislocation motion. Notably, local cross-slip events are identified in the Nb$_{Al}$ and antisite models, as evidenced by lateral displacement signatures along the $y$-direction (Figure 6d). In contrast, such cross-slip behavior was absent in the Nb$_{Ti}$ model, indicating distinct defect-dislocation interactions depending on the substitution site.

The Peierls stress results of edge dislocations for different defect models are shown in Figure S10. For the Nb-free model, the Peierls stress of the edge dislocation is ~25 MPa, which is substantially lower than that of the screw dislocation (~170 MPa). This indicates that edge dislocations move much more easily than screw dislocations in γ-TiAl. After introducing defects, the Peierls stress of edge dislocations increases significantly: ~82 MPa for the 8 at.% Nb$_{Ti}$ model and ~205 MPa for the 8 at.% Nb$_{Al}$ model (Figure S10), demonstrating a stronger strengthening effect when Nb substitutes for Al atoms. Moreover, antisite defects yield higher Peierls stress than substitutional defects, consistent with the behavior observed for screw dislocations (Figure 6b). Overall, both screw and edge dislocations exhibit a similar strengthening trend: Al$_{Ti}$ > Ti$_{Al}$ > Nb$_{Al}$ > Nb$_{Ti}$ > Nb-free. These results confirm that Nb substitution at Ti sites does enhance the strength of γ-TiAl, although its strengthening effect is weaker than that of Nb$_{Al}$ and antisites. Thus, in the absence of microstructural effects, Ti-rich high-Nb TiAl alloys, which usually contain larger populations of Ti$_{Al}$ and Nb$_{Al}$ defects (Figure S7), are expected to display a notably stronger solid-solution strengthening response than Al-rich alloys. This may help explain the experimentally observed[11] trend that Nb strengthening is more evident in Ti-rich than in Al-rich alloys.

***Implications for Defect and Composition Engineering in Nb-Doped TiAl and Beyond***

The mechanistic insights uncovered in this study offer valuable guidance for designing Nb-doped γ-TiAl alloys with superior mechanical properties through deliberate manipulation of defect chemistry and compositional tuning. Our simulations reveal that Nb not only substitutes preferentially at Ti sites, but also partially occupies Al sites at higher concentrations, giving rise to antisite defects and complex local environments. This finding challenges the conventional assumption that Nb's role is limited to solid-solution strengthening through Ti-site substitution and instead highlights a broader range of defect-mediated interactions. Importantly, the presence of Nb$_{Al}$ and Ti$_{Al}$ antisite defects significantly reduces stacking fault energies, thereby promoting deformation twinning and the formation of high-density nanotwin structures[9]. These twins act as potent barriers to dislocation motion, enabling simultaneous enhancement of strength and ductility—a desirable yet often conflicting goal in intermetallic alloy design. At the same time, the increase in Peierls stress associated with Nb substitution and antisite defects confirms a robust solid-solution strengthening contribution. The concurrent presence of SRO, reduced SFEs, and elevated Peierls stress suggests that careful control over site occupancy and defect populations can yield a synergistic combination of mechanical properties.

These results point to several useful design principles for high-performance TiAl-based alloys: First, controlled defect populations via targeted Nb content and heat treatment can balance solid-solution hardening with twin-induced plasticity. Second, adjusting short-range order by controlling local chemical environments may serve as another lever to tune strength and ductility. Finally, site-selective doping strategies can be exploited to tailor stacking fault energies and dislocation behavior, a strategy that should be applicable to other ordered intermetallics beyond $\gamma$-TiAl. The methodology demonstrated here, combining high-fidelity machine learning potentials with large-scale hybrid Monte Carlo/molecular dynamics simulations, can be extended to other complex alloys, where defect interactions govern mechanical behavior. In particular, it provides a practical framework for exploring solute-site competition, defect complexes, and the interplay between chemical order and plasticity in next-generation high-temperature structural alloys.

**CONCLUSIONS**

In this work, we use a high-fidelity NNP model to clarify the role of Nb in defect-mediated strengthening and plasticity of $\gamma$-TiAl. MCMD simulations show that Nb preferentially substitutes Ti sites and forms SRO with surrounding Al atoms, while a notable fraction also occupies Al sites at higher alloying concentrations, accompanied by antisite formation. GSFE calculations show that $Nb_{Ti}$ increases SFEs, whereas $Nb_{Al}$ and antisite defects markedly reduce them. Both $Nb_{Al}$ and $Ti_{Al}$ defects promote deformation twinning structures. These twins improve ductility by offering additional deformation modes and increase strength by obstructing dislocation motion, consistent with experimentally observed SFE reductions. Peierls-stress results further show that Nb substitution and antisite defects significantly increase the critical stress for dislocation glide, confirming strong solid-solution strengthening. MCMD simulations also show that Ti-rich high-Nb alloys contain larger populations of $Nb_{Al}$ and $Ti_{Al}$ defects, which strengthen more effectively than $Nb_{Ti}$ defects and help explain why Ti-rich high-Nb TiAl alloys exhibit higher strength than their Al-rich counterparts. Together, these findings suggest that the exceptional mechanical performance of high-Nb $\gamma$-TiAl arises from the synergetic interplay of solid-solution strengthening, deformation twinning-mediated plasticity, and chemical short-range ordering. This study provides a unified, atomistically-informed explanation for the Nb-induced enhancement of both strength and ductility in γ-TiAl and offers practical guidelines for defect and composition engineering in advanced alloy systems.

## DATA AND CODE AVAILABILITY

All data are provided in the article and Supplemental information or is available from the corresponding authors upon request.

## SUPPLEMENTAL INFORMATION

Supplemental information can be found online at https://doi.org/10.1016/j.acta.xxxx.

## ACKNOWLEDGMENTS

This work was supported by the AI Singapore Grand Challenge in AI for Materials Discovery Funding Scheme (AISG2-GC-2023-010), National Key Research and Development Program of China (2019YFA0705400), National Natural Science Foundation of China (11772153, 22073048), the Natural Science Foundation of Jiangsu Province (BK20190018), the Research Fund of State Key Laboratory of Mechanics and Control of Mechanical Structures (MCMS-E-0420K01), the Free Exploration Science Foundation (2023Z053052001), and a Project by the Priority Academic Program Development of Jiangsu Higher Education Institutions.

## AUTHOR CONTRIBUTIONS

Z.Q. Z. designed the project, performed the theoretical calculations and wrote the original manuscript. Y.-W. Z and Z.H. Z. contributed to the idea generation, data analysis, writing, and funding support. W.L. G., and S. O contributed to the data analysis and writing. M.L. Y., J.P. D., S.Y. S., K.P. O.Y., L.L. C., X.K. C helped with discussion of the results. All the authors discussed the results and commented on the manuscript.

## DECLARATION OF INTERESTS

The authors declare no competing interests.

using atomistic simulations. *Materials & Design* **237**, 112596 (2024).